\documentclass[twocolumn,showpacs,amsmath,amssymb,prl,superscriptaddress,floatfix]{revtex4}

\usepackage{graphicx}
\usepackage{dcolumn}
\usepackage{bm}
\newcommand{\bg}[1]{\mbox{\boldmath$#1$}} 
\usepackage{xcolor}
\definecolor{red}{rgb}{0.7,0,0}
\definecolor{green}{rgb}{0.,0.35,0.}
\definecolor{blue}{rgb}{0.2,0.2,0.7} 
\definecolor{black}{rgb}{0.15,0.15,.15}

\begin{document}

\title{Strongly correlated gases of Rydberg-dressed atoms: quantum and classical dynamics}

\author{G. Pupillo}
\affiliation{Institute for
Theoretical Physics, University of Innsbruck, and Institute for
Quantum Optics and Quantum Information of the Austrian Academy of
Sciences, Innsbruck, Austria}
\author{A. Micheli}
\affiliation{Institute for
Theoretical Physics, University of Innsbruck, and Institute for
Quantum Optics and Quantum Information of the Austrian Academy of
Sciences, Innsbruck, Austria}
\author{M. Boninsegni}
\affiliation{Department of Physics, University of Alberta, Edmonton, Alberta, Canada T6G 2J1}
\affiliation{Institute for
Theoretical Physics, University of Innsbruck, and Institute for
Quantum Optics and Quantum Information of the Austrian Academy of
Sciences, Innsbruck, Austria}
\author{I. Lesanovsky}
\affiliation{School of Physics and Astronomy, University of Nottingham, Nottingham, UK}
\author{P. Zoller}
\affiliation{Institute for Theoretical Physics, University of
Innsbruck, and Institute for Quantum Optics and Quantum Information
of the Austrian Academy of Sciences, Innsbruck, Austria}
\begin{abstract}
We discuss techniques to generate long-range interactions in a gas of groundstate alkali atoms, by weakly admixing excited Rydberg states with laser light. This provides a tool to engineer strongly correlated phases with reduced decoherence from inelastic collisions and spontaneous emission. As an illustration, we discuss the quantum phases of dressed atoms with dipole-dipole interactions confined in a harmonic potential, as relevant to experiments. We show that residual spontaneous emission from the Rydberg state acts as a heating mechanism, leading to a quantum-classical crossover.
\end{abstract} \maketitle

There is currently significant interest in the physics of dipolar quantum degenerate gases~\cite{Baranov}. Strong long-range dipole-dipole interactions promise the realization of novel many-body phases in neutral gases, such as self-assembled crystals~\cite{Micheli07}, topological superfluids and quantum phases with hidden topological order~\cite{Cooper09}.  The regime of strong dipolar couplings is easily accessible with interacting {\em electric} dipole moments, as realized in particular in quantum gases with polar molecules prepared in their rovibrational ground state~\cite{LahayeCarr}. In contrast, quantum gases of ground state atoms typically interact via the much smaller {\em magnetic} interactions~\cite{ChromiumExp}. The question, therefore, is to what extent this regime of strong dipolar interactions can also be realized with present atomic gases experiments with alkali atoms. Here we propose and investigate a setup where the huge electric dipole moments $d \sim n^2$~\cite{GallagherBook,RydbergExp,Santos} of atomic Rydberg states with principal quantum number $n$ are weakly admixed to the atomic ground state, thus providing an atomic gas of interacting effective electric dipoles comparable to the case of polar molecules.  A central question, and the main difference to the molecular case, is decoherence and heating mechanisms associated with spontaneous emission from Rydberg states, and possible inelastic collisions. Below we show that (i) dipolar crystals can be realized with Rydberg-dressed atoms confined to two dimensions, with negligible collisional losses, and (ii) spontaneous emission $\gamma_{\rm r}$ provides an intrinsic heating mechanism, allowing the study of thermalization phenomena with full tunability of system parameters. We study the associated crystal melting with molecular dynamics calculations.

\begin{figure}
\includegraphics[width=0.9\columnwidth]{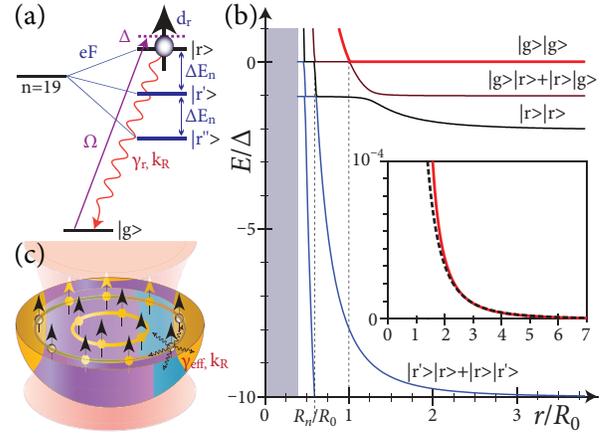}
\caption{(color online)(a) Sketch of the energy levels constituting the Stark fan of an alkali metal atom exposed to an electric field. $|g\rangle$ and $|r\rangle$ are coupled by a laser with Rabi frequency $\Omega$ and (blue) detuning $\Delta$. $\gamma_{\rm r}$ is the spontaneous emission from $|r\rangle$, with momentum $k_R$ of the photon recoil. (b) BO potentials in the ($x-y$)-plane in the dressed picture. $R_0$ is the resonant Condon point. The effective interaction potential $V_{\rm int}^{\rm 3D}(r)$ is the higher-energy curve (red). Level crossings occur for $R_{\rm n}<R_0$. Inset: blow-up of $V_{\rm int}^{\rm 3D}(r)$ (continuous line) compared to $1/r^3$ (dashed line). (c) Sketch of experimental setup: Rydberg-dressed atoms are confined to 2D by a strong confining laser beam, with dipoles polarized perpendicular to the plane. An effective spontaneous emission with rate $\gamma_{\rm eff}$ provides for an intrinsic heating mechanism (see text).}\label{fig:fig1}
\end{figure}

The setup we have in mind is illustrated in Fig.~\ref{fig:fig1}: we propose to weakly couple with laser light the groundstate $|g\rangle$ of each atom to a Stark-split Rydberg state $|r\rangle$ with large dipole moment $d_0$, in the kDebye. For large enough detuning $\Delta$ from resonance and interparticle distances, interactions are of the dipole-dipole type $V_{\rm int}^{\rm 3D} \propto (\Omega/\Delta)^4 d_0^2/r^3$, with $\Omega$ the Rabi frequency. By confining the particles to a 2D plane using an optical field, the effective in-plane interactions  $V_{\rm int}^{\rm 2D}$ are then purely repulsive, with negligible collisional losses. This opens the way to the study of the many-body phases of 2D dipoles in these systems. As an illustration, we show the existence of mesoscopic supersolids and crystals with Rydberg-dressed atoms under realistic conditions of in-plane harmonic confinement, using exact quantum Monte-Carlo simulations. Residual spontaneous emission $\gamma_{\rm eff} \sim (\Omega/\Delta)^2\gamma_{\rm r}$ from the Rydberg state introduces an intrinsic heating mechanism, driving the quantum phases into the classical regime. We study the quantum/classical crossover by means of molecular dynamics simulations, and show the emergence of a dynamical thermalization timescale in these systems.

\textit{Effective interaction potentials} with suppressed decoherence from atomic collisions and spontaneous emission are obtained as follows. In the presence of a homogeneous electric field (field strength $F$, oriented along $\mathbf{e}_z$) the energy levels of an alkali atom show the well-known Stark structure~\cite{GallagherBook}. We are interested here in high angular momentum states which show a linear Stark effect. 
The energy of these states is well approximated by $\epsilon_{nn_1n_2}= (3/2)e a_0 n (n_1-n_2) F$, with $n_1$ and $n_2$ parabolic quantum numbers, $a_0$ the Bohr radius and $e$ the electron charge. The highest-energy state $|r\rangle$ of a given manifold $n$ ($n_1=n-1$, $n_2=0$) 
is energetically separated from the next (lower lying) state in the manifold by $\Delta E_{n}=e (3/2) a_0 n\,F$, while adjacent manifolds remain well separated by an energy $\Delta E_{\rm IT} \gg \Delta E_{n}$ for field strengths smaller than $F_\mathrm{IT}\approx \frac{1}{3\, n^5} \frac{m^2e^5}{128 \pi^3 \epsilon_0^3 \hbar^4}$ (Inglis-Teller limit). 
In our scheme, Fig.~\ref{fig:fig1}, 
each atom is effectively reduced to a two-level system. Within the rotating-wave approximation and with a product basis $\left\{\left|r\right>_i\!\left|r\right>_j,\left|r\right>_i\!\left|g\right>_j,\left|g\right>_i\!\left|r\right>_j,\left|g\right>_i\!\left|g\right>_j\right\}$
the Hamiltonian governing the internal dynamics of two atoms is given by
\begin{eqnarray}
  H_{ij}=\left(
       \begin{array}{cccc}
         -2\Delta+v_{ij} & \Omega & \Omega & 0 \\
         \Omega & -\Delta & 0 & \Omega \\
         \Omega & 0 & -\Delta & \Omega \\
         0 & \Omega & \Omega & 0 \\
       \end{array}
     \right),
\end{eqnarray}
with $v_{ij}=D(1-3\cos\theta_{ij})\,|\mathbf{R}_i-\mathbf{R}_j|^{-3} \equiv D \, \nu_{ij} \,R_{ij}^{-3}$, the dipole-dipole interaction between two atoms at position $\mathbf{R}_i(\mathbf{R}_j)$,  and $D=d_0^2/(4\pi\epsilon_0)$. $\theta_{ij}$ is the angle between the vector $\mathbf{R}_i-\mathbf{R}_j$ and the dipole moment aligned parallel to the $z$-axis. 
The interaction potential which is experienced by the dressed ground state atoms is given by the Born-Oppenheimer (BO) energy surface that adiabatically connects to the energy of the product state $\left|g\right>_i\!\left|g\right>_j$ as $\Delta\rightarrow\infty$. Here, we focus on the weak driving limit $\Omega < \Delta$. For two atoms approaching each other and blue detuning $\Delta > 0$, the groundstate BO potential will be approximately given by $V_{\rm int}^{\rm 3D}(R_{ij},\theta_{ij}) \simeq D \nu_{ij}/R_{ij}^3$ for $R_{ij}<R_0$, with $R_0\simeq (D/\hbar \Delta)^{1/3}$ a resonant Condon point with typical values in the hundreds of nm, Fig.~\ref{fig:fig1}(b). Diabatic crossings with different potential surfaces of the same and different $n$-manifolds leading to collisional {\it two-body} losses will occur in this region $R_{ij}<R_0$, at distances of order $R_{n}\simeq (D/\Delta E_n)^{1/3}$ and $R_{IT}\simeq (D/\Delta E_{IT})^{1/3}$, respectively, with $R_0 > R_n\gg R_{IT}$, reminiscent of blue-shielding techniques~\cite{Weiner99,Micheli07}.

For $R_{ij} \gtrsim R_0$, two-body diabatic losses are absent and  $V_{\rm int}^{\rm 3D}(R_{ij},\theta_{ij}) \simeq (\Omega/\Delta)^4 D \nu_{ij}/R_{ij}^3$. In this work we will focus on this parameter regime.
This has two additional advantages: (i) spontaneous emission rates are strongly reduced to values $\gamma_{\rm eff}$ 
and (ii) effective interactions are reduced to values compatible with trapping of atoms with optical fields (e.g., optical lattice), and confinement to low-dimensional geometries. 

Collisional losses for $R_{ij}>R_0$ are linked to population of the attractive part of the dipole-dipole interaction. Sampling of this attractive part can be suppressed by confining atoms to 2D using a tight (optical) trapping along $z$, with harmonic frequency $\hbar \omega_\perp = \hbar^2/ m a_{\perp}^2 > V_{\rm int}^{\rm 3D}(R)$, with $m$ the atomic mass, and $a_{\perp}$ the transverse harmonic oscillator length. The BO potential is then purely repulsive in 3D: there is an energy barrier between the long-distance repulsion and the attractive short-distance regime. Residual losses are linked to tunneling below this energy barrier, and can be computed semiclassicaly as $\Gamma_{\rm coll} = \omega_{\rm att} \exp[-c( \Omega^4 D m/ \hbar^2 \Delta^4 a_\perp)^{2/5}]$, with $c$ a constant of order unity, and $\omega_{\rm att}$ the attempt frequency, of order of the average kinetic energy in the gas. For $R_{ij} > R_0$ the effective dynamics is purely 2D, with interactions $V_{\rm int}^{\rm 2D}= (\Omega/\Delta)^4 D/\rho^3 \equiv \tilde D/\rho^3$, with $\rho = |{\bg \rho}|$, and ${\bg \rho}$ a vector in the ($x-y$)-plane.

\begin{figure}
\includegraphics[width=0.8\columnwidth]{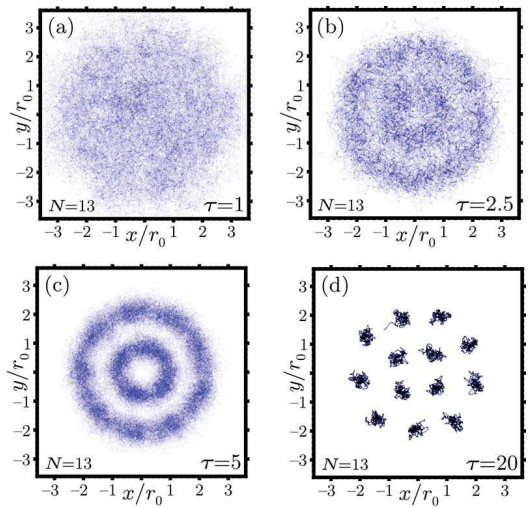}
\caption{(color online) (a-d) Monte Carlo snapshots of the density of particles in all mesoscopic phases for $N=13$ dipoles, as a function of the effective mass $\tau$. (a) superfluid; (b) supersolid; (c) ring-like crystals; (d) classical crystal.}\label{fig:fig2}
\end{figure}

{\it Mesoscopic crystals with Rydberg-dressed atoms:} We consider a setup where $N$ bosonic dressed atoms are confined to a 2D plane by applying a strong transverse trapping field~\cite{Micheli07}, e.g a 1D optical lattice, and are aligned perpendicular to the plane [Fig.~\ref{fig:fig1}(c)]. We assume an additional in-plane parabolic trap with frequency $\omega$, as realized in experiments by a magnetic dipole trap, or a single site of a large spacing optical lattice.  Defining length and energy scales $r_0=(\tilde D/m\omega^2)^{1/5}$ and $\tilde \epsilon=m\omega^2r_0^2=\tilde D/r_0^3=(m^3\omega^6 \tilde D^2)^{1/5}$, respectively, the 2D Hamiltonian in dimensionless form reads
\begin{eqnarray}\label{eq:eqHamRes}
\frac{H}{\tilde \epsilon}=\sum_{i=1}^{N} \left[-\frac{1}{2 \tau^2}\frac{\partial^2}{\partial {\bg \rho}_i^2}+\frac{1}{2}{\bg \rho}_i^2\right]
    + \sum_{i>j}\frac{1}{|{\bg \rho}_i-{\bg \rho}_j|^3},
\end{eqnarray}
where $ \tau \equiv \tilde \epsilon / \hbar \omega=\left(r_0/ \ell\right)^2 = \left(m \tilde D/ \hbar^2 \ell\right)^{2/5}$ characterizes the strength of the dipole-dipole interactions in the trap with $\ell = \sqrt{\hbar/m\omega}$ the harmonic oscillator length. Equation~\eqref{eq:eqHamRes} shows that $\tau$ plays the role of an {\em effective mass} representing a control parameter, which can be increased by increasing the strength of dipole-dipole interactions, or by compressing the trap.

\begin{figure}
\includegraphics[width=0.8\columnwidth]{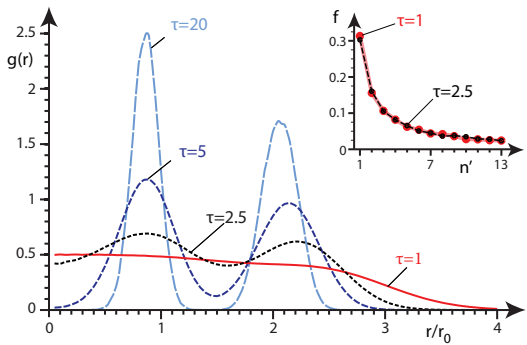}
\caption{(color online) Radial density profiles $g(r)$ for the cases of Fig.~\ref{fig:fig2}. Inset: statistics of computed particle exchanges $f$ as a function of the number of particles $n'$ participating to the exchange for the cases of finite superfluid fraction $\tau=1$ and 2.5. In a superfluid, $f$ is finite for all $n' \leq N$. The supersolid phase with $\tau=2.5$ has finite density modulation (figure) and the same $f$-distribution of $\tau=1$ (inset).}\label{fig:fig2}
\end{figure}
For small $\tau \lesssim 1$, we expect the kinetic energy to dominate, and the cluster to be in a weakly interacting superfluid (SF) phase, while for $\tau \gg 1$ the kinetic energy becomes negligible, and the system ground state should resemble the classical crystalline (CC) configuration obtained by minimizing the last two terms of Eq.~\eqref{eq:eqHamRes}. We obtain an estimate of the critical $\tau_{\rm c}$ for the crossover from the superfluid to the crystal by noting that for a homogeneous system the superfluid-crystal transition occurs at $r_{\rm QM}= D m/\hbar^2 a= 18 \pm 4$~\cite{Micheli07,Astrakharchik07}, where $r_{\rm QM}$ represents the ratio of the dipolar interactions $D/a^3$ to the kinetic energy $\hbar^2/ma^2$ with $a$ the mean interparticle distance. By rewriting $\tau_{\rm c} = (r_{\rm QM} a / \ell)^{2/5}$, and approximating $a \sim \ell$, we obtain the prediction $\tau_{\rm c} \simeq 3$, which is essentially $N$-{\it independent}.

We determine the generic zero-temperature phase diagram for trapped dipolar atoms by means of Quantum Monte Carlo simulations based on the continuous-space Worm Algorithm~\cite{worm}. We find that for mesoscopic clusters with $N \lesssim 40$ four different quantum phases exist: (i) a SF for $\tau \ll \tau_c$, (ii) a mesoscopic supersolid (MS) for $\tau \lesssim \tau_c$, (iii) a ring-shaped crystal (RC) for $\tau \gtrsim \tau_c$, and (iv) a CC for $\tau \gg \tau_c$. These phases are distinguished by measuring the superfluid fraction $\rho_{\rm s}$ and the radial density profile $g(r)$: the SF and MS phases are superfluid with $\rho_{\rm s}= 1$, while the RC and CC phases have $\rho_{\rm s}= 0$; the SF has a flat, featureless $g(r)$, while the MS, RC and CC phases have significant density modulations. In the crystalline RC phase, particles are arranged in concentric rings with a {\it fixed} number of particles per ring. For small enough temperatures, these rings are free to rotate independently of each other. In the crystalline CC phase, atoms are arranged in the classical crystal configuration, at fixed relative positions. Mesoscopic crystals have been also found in excitonic materials in Refs.\cite{Lozovik04}.

In the following we provide example results for $N=13$ atoms, which we found to display all general features of mesoscopic clusters with $N \lesssim 40$. We find that for $N \gg 40$ the system resembles the homogeneous situation, with a sharp crossover between SF and CC phases around $\tau_c$.

Panels (a-d) in Fig.~\ref{fig:fig2} are snapshots of the particle density for $N=13$ and $1 \leq \tau \leq 20$ in the four mesoscopic phases described above. These results correspond to a low enough temperature $T \ll \tilde \epsilon$, that they can be regarded as groundstate estimates. Panel (a) shows a SF for $\tau =1$ with a flat density profile (see also $g(r)$ in Fig.~\ref{fig:fig2}), where the various particle probability clouds overlap. This overlap is directly connected to superfluidity~\cite{Sindzingre89}, and consistently we here measure $\rho_{\rm s}=1$. The emergence of a MS phase for $\tau = 2.5 \lesssim \tau_{\rm c}$ is signaled by a distinguishable density modulation in Fig.~\ref{fig:fig1}(b) (see also Fig.~\ref{fig:fig2}), combined to a measured $\rho_{\rm s}=1$.
We find that the superfluid properties are completely unaltered by the increased strength of interactions with respect to the case $\tau=1$. This is quantified in the inset of Fig.~\ref{fig:fig2} by measuring the statistics of exchange cycles $f$ in a many-particle path~\cite{Sindzingre89}: in a superfluid, the probability $f$ that $n'$ particles exchange is finite for all $n' \leq N$, while in a crystal is approximately zero. In the inset of Fig.~\ref{fig:fig2} it is shown that $f$ is finite and equal for $\tau=1$ and 2.5, computed at corresponding low temperatures (i.e., same fractions of $\tilde \epsilon$). %

For $\tau \gtrsim \tau_{\rm c}$ [panels (c-d)] we observe crystallization of the atomic cloud, with $\rho_{\rm s}=0$ and finite density modulations, consistently with the estimates above.
In particular, for $\tau=5$ we obtain a RC phase, with concentric, independent rings~\cite{note}, while for $\tau=20$ particles are arranged in the CC configuration for $N=13$, with 4 particles at the center, and 9 outside. This classical crystal, characterized by large peaks in $g(r)$ in Fig.~\ref{fig:fig2}, is found to be the groundstate configuration for all $\tau\gtrsim 20$.

\begin{figure}[b]
\includegraphics[width=0.9\columnwidth]{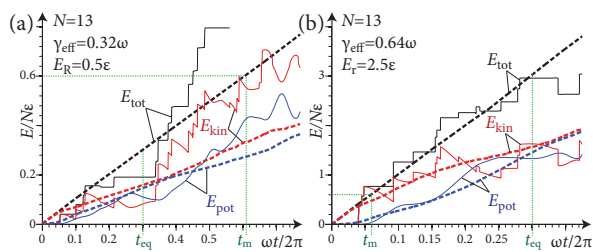}
\caption{(color online) Molecular dynamics (MD) simulation of crystal heating. Black, red and blue lines are total energy $E_{\rm tot}$, kinetic $E_{\rm kin}$ and potential $E_{\rm pot}$ energies vs. time $t$, respectively. Thin continuous lines: single MD trajectories. Thick dashed lines: averages over many MD trajectories. $t_{\rm eq}$ and $t_{\rm m}$ are the equilibration and melting time, respectively.}\label{fig:fig4}
\end{figure}

{\em Effective heating}: The residual effective single-particle spontaneous emission rate is $\gamma_{\rm eff}\sim(\Omega/\Delta)^2\gamma_{\rm r}$. In this work, we focus on the case where after the spontaneous emission the atom is found in $|g\rangle$, and then instantaneously re-dressed, which is the relevant situation for $n \lesssim 30$~\cite{MicheliUnpublished}.
Spontaneous emission results in the release of a photon, with momentum $\hbar k$ and recoil energy $E_R$, in the tens of kHz. For dipoles parallel to $F$, emission will be preferentially along $z$, and $E_R$ will be absorbed by the confining lattice potential with $\hbar \omega_{\perp} > E_R$. However, any finite component of the momentum $\hbar k_{xy}$ of the emitted photon in the ($x-y$)-plane will result in a random "kick" of the atoms, which will ultimately translate into an effective intrinsic heating rate for the many-body system. We conservatively estimate the crystalline phases ($\tau\gg1$) to be observable up to an average ``melting'' time $t_{\rm m}\propto  T_{\rm M}/(\gamma_{\rm eff}E_{R})$, with $T_{\rm M}$ the (classical) melting temperature in the trap.
We computed $T_{\rm M}$ for all $N<30$ by means of molecular dynamics simulations for the 2D system in the classical limit ($\tau\rightarrow\infty$)~\cite{Kalia81}. By monitoring the behavior of the specific heat and density-density correlations, we determine the value $T_{\rm M} = 0.4(2) \tilde \epsilon$. This can result in characteristic lifetimes of order of several milliseconds. For example, for $^{87}$Rb atoms coupled to the Rydberg-state ($n=20$, $n_1=19$, $n_2=0$) with $d_0\approx1.45{\rm kDebye}$, $\gamma_r \sim 100 $kHz, $E_R \simeq 25$kHz, via a laser with $\Omega/2 \pi=100{\rm MHz}$, $\Delta/2\pi=1{\rm GHz}$ and a DC field $F=25$kV/m, for $\omega/2 \pi = 5$kHz we obtain $r_{\rm 0} \simeq 680$nm, $\tau \simeq 20$, $\tilde \epsilon \simeq 4 \, E_R$,  and $t_{\rm m} \gtrsim 5$ms.

We further investigated the melting dynamics. Spontaneous emission was simulated by applying random in-plane kicks to each particle with momentum $\hbar k_{xy} = (2mE_{Rxy})^{1/2}$, with fixed $E_{Rxy}\lesssim E_R$, at an average rate $\gamma_{\rm eff}$. Panels (a) and (b) in Fig.~\ref{fig:fig4} show our results for $E_{Rxy}=0.5 \tilde \epsilon$ and 2.5 $\tilde \epsilon$ as a function of time $t$,
respectively, for $N=13$ particles [see Fig.~\ref{fig:fig2}]. Thin lines are
single classical-dynamics trajectories, while thick dashed lines are averages over many trajectories. We find that the average energy $E_{\rm tot}$ increases linearly with time $t$, as expected. For the smaller $E_{Rxy}$ of panel (a), we find that the crystal melts roughly at the expected time $t_{\rm m}$ (corresponding to several spontaneous emission processes), while in panel (b) the observed melting time $t_{\rm eq}$ is significantly longer than $t_{\rm m}$ (roughly the time of one spontaneous emission). This prolonged stability is a dynamical process: classical melting can occur only after the system has reached thermal equilibrium. The latter corresponds to the time $t_{\rm eq}$ at which the average potential energy $E_{\rm pot}$ and kinetic energy $E_{\rm kin}$ become equal and equipartition of energy applies. $t_{\rm eq}$, and thus the crystal lifetime, can be (much) longer than $t_{\rm m}$.

Since interparticle distances can be of the order of a $\mu$m or more, the {\it spatial structure} of the crystalline phases can be imaged using, e.g., tightly focused beams.

The authors thank H.P.~B\"uchler and F.~Ferlaino for discussions. This
work was supported by IQOQI, the Austrian FWF, the US through MURI and EOARD, the EU through STREP FP7-ICT-2007-C project NAME-QUAM, the Canadian NSERC through G121210893.


\begin{thebibliography}{40}

\bibitem{Baranov}T. Lahaye {\it et al.}, Rep. Prog. Phys. {\bf 72}, 126401 (2009); R.V. Krems, W.C. Stwalley and B. Friedrich, \textit{Cold Molecules: Theory Experiment, Applications} (CRC Press, 2009).

\bibitem{Micheli07}H. P. B\"uchler \textit{et al.}, Phys. Rev. Lett. \textbf{98}, 060404 (2007);
A. Micheli \textit{et al.}, Phys. Rev. A \textbf{76}, 043604 (2007).

\bibitem{Cooper09}N.R. Cooper and G. Shlyapnikov, Phys. Rev. Lett. \textbf{103}, 155302 (2009); E. G. Dalla Torre, E. Berg, and E. Altman, \textit{ibid.} {\bf 97}, 260401 (2006).

\bibitem{LahayeCarr} L.D. Carr {\it et al.}, New J. Phys. {\bf 11}, 055049 (2009); M. Baranov, Phys. Rep. {\bf 464}, 71 (2008).


\bibitem{ChromiumExp}
{T. Lahaye \textit{et al.}, Nature \textbf{448}, 672-675 (2007); M Fattori \textit{et al.}, Phys. Rev. Lett. \textbf{101}, 190405 (2008). }

\bibitem{GallagherBook} 
{T. F. Gallagher, \textit{Rydberg Atoms}, Cambridge University Press (1994); M.L. Zimmerman {\it et al.}, Phys. Rev. A \textbf{20}, 2251 (1979).}


\bibitem{RydbergExp}
{K. Singer \textit{et al.}, Phys. Rev. Lett. \textbf{93}, 163001 (2004); D. Tong \textit{et al.}, {\it ibid.} {\bf 93}, 063001 (2004); T. Vogt \textit{et al.}, {\it ibid.} {\bf 97}, 083003 (2006); R. Heidemann \textit{et al.}, {\it ibid.} \textbf{100}, 033601 (2008); U. Raitzsch \textit{et al.},  {\it ibid.} \textbf{100}, 013002 (2008); H. Weimer \textit{et al.},  {\it ibid.} \textbf{101}, 250601 (2008); C. Roux \textit{et al.}, Eur. Phys. Lett. {\bf 81}, 56004 (2008); A. Ga\"etan \textit{et al.}, Nature Phys. \textbf{5}, 115 (2009); E. Urban \textit{et al.},  {\it ibid.} \textbf{5}, 110 (2009); H. Schempp \textit{et al.}, arXiv:0912.4099. }

\bibitem{Santos}{L. Santos \textit{et al.}, Phys. Rev. Lett. {\bf 85}, 1791 (2000); M.D. Lukin \textit{et al.},  {\it ibid.} \textbf{87}, 037901 (2001); M. Saffman and K. M$\o$lmer, {\it ibid.} \textbf{102}, 240502 (2009); T. Pohl, E. Demler, M. D. Lukin, arXiv:0911.1427.}

\bibitem{Weiner99}{J.~Weiner, V. S. Bagnato, S. Zilio, and P. S. Julienne, Rev.~Mod.~Phys.~\textbf{71}, 1 (1999).}

\bibitem{Astrakharchik07} 
{G. E. Astrakharchik \textit{et al.}, Phys. Rev. Lett. \textbf{98}, 060405 (2007).}


\bibitem{worm} M. Boninsegni, N. V. Prokof'ev and B. V. Svistunov, Phys. Rev. Lett. {\bf 96}, 070601 (2006) and Phys. Rev. E {\bf 74}, 036701 (2006).

\bibitem{Lozovik04} A.I. Belousov and Yu.E. Lozovik, Eur. Phys. J. D \textbf{8}, 251 (2000); Yu.E. Lozovik, S. Yu. Volkov, and M. Willander, JETP Lett. {\bf 79}, 473, (2004); P. Ludwig \textit{et al.}, New J. Phys. \textbf{10}, 083031 (2008). For Coulomb clusters: V.M. Bedanov and F.M. Peeters, Phys. Rev. B {\bf 49}, 2667 (1994); A. Filinov \textit{et al.}, {\it ibid.} {\bf 77}, 214527 (2008).




\bibitem{Sindzingre89}P. Sindzingre, M. L. Klein and D. M. Ceperley, Phys. Rev. Lett. {\bf 63}, 1601 (1989).

\bibitem{note} Finite clusters can rotate at a temperature $ T \lesssim \hbar^2/I$, $I$ being the classical moment of inertia.  We have observed this for $\tau=5$: $f$ shows two sharp peaks for $n=4$ and 9 (inner and outer rings, respectively), other cycles being absent. However, this is not a ``superfluid".

\bibitem{MicheliUnpublished}
Choosing larger $n$ can make lifetimes  longer, since $\gamma_{\rm r} \sim 1/n^3$. However, then it is not possible to neglect processes where the atom is found in an excited state $|r'\rangle \neq |r\rangle$ after the spontaneous emission; S.~Thwaite \textit{et al.}, unpublished.



\bibitem{Kalia81} 
{R. K. Kalia and P. Vashishta, J. Phys. C \textbf{14}, L643 (1981).}



\end{thebibliography}
\end{document}